# Algae-like Artificial Organic Photo-tactic Micro-swimmers

Somnath Koley[1,2 *], and Karuna Kar Nanda[1,2,3]

[1]*Institute of Physics Bhubaneswar, Odisha-751005, India*
[2] *Homi Bhabha National Institute, Mumbai, Maharashtra-400094, India*
[3]*Materials Research Centre, Indian Institute of Science, Bangalore-560012, India*
[*]Corresponding Author, Email: somnath.koley@iopb.res.in

**Abstract:** Phototaxis is a light driven self-locomotion of mass and a common phenomenon in motile organisms with varieties of motility such as in bacteria, algae, etc. In naturally occurring organisms, mechanical force is generated utilizing their metabolic energy to propel and swim in presence of light, performing important bio-cemical reactions. These observations spurred preparation of artificial counterparts for controlling the motion for targeted applications and creation of a library of microswimmers exhibiting interesting characteristics. Herein, we report a new class of microswimmers that exhibit captivating and complicated micro-swimming mediated colony formation properties resembling green algae. A facile pyrolysis reaction is explored leading to homogeneous organic nanostructures forming patterned self-assemblies among themselves. A delicate balance of lyophilicity and surface charge of the assembled colloidal nano-carbons forms interesting architectures such as dynamic colonies, thallus like patterning, cilia like micro-arms. All these complex characteristics have been studied with visualized experiments, optical spectroscopy and microscopy, and electron microscopy. In presence of weak light intensity, both positive and negative phototaxis are seen moving the microswimmers propelling towards and away from the light respectively. During swimming helical motion and electrostatic interactions of colloidal microswimmers with neighbouring assemblies are observed. The nature of assembly formation is found to be fractal and can be disintegrated using strong light. Strong exposure stimulates predominately fast negative phototaxis leading to directional propulsion along the light path. All these algae life-like behaviour of the colloidal carbonaceous lyophilic sols are stable in alcohol and can be reversibly discontinued with water due to superhydrophilicity. We therefore introduce a class of active and photo-responsive colloidal chemical architecture with complex life-like behaviour termed as "alco-algae". Our discovery may entice studies for creation of a diverse programmable micro-swimmers for microscopic understanding and manipulating collective effect of the assemblies for bio-mimetic and catalytic applications.

Mimicking organ-like phenomena in artificial functional systems have been of immense interests to meet scientific curiosity and development of a library of bio-inspired synthetic counterparts of important fundamental organisms[1-4]. The artificial architectures consisting of lab grown complex chemical structures, do not replicate essentially all of the mechanisms of life, but essentially exhibit a key behaviour analogous to the naturally occurring lively organisms in presence of a stimuli[5-14]. The stimuli must activate the chemical body and the body transforms the absorbed energy into bio-chemical, electrical, and mechanical energies to perform important processes such as photosynthesis, signalling, and directed motion respectively. If we take the example of naturally occurring green algae, it is an interesting system that can perform multitude of energy harvesting processes such as photo-synthesis, phototactic locomotion via micro-swimming, and colony formation [15-18]. Micro-swimming in algae is an essential process for energy harvesting because using positive phototaxis the algae bodies or colonies moves towards the direction of light to gain enough energy for photosynthesis and using negative phototaxis, they move away from the light source to avoid exposure induced damage. The swimming gaits are interestingly complex in algae and vary in different species. The beating of swimming arms called flagella and cilia, in various ways such as trot, pronk and gallop of both transverse and rotary-types [19]. This inspires to understand these complex behaviour and factor affecting them by studying model biological systems, and creating different type of chemical microswimmers responding in an optical landscape [5, 15]. Naturally occurring motile microorganisms move using their metabolic energy. Artificial microswimmers are important class of active matter possessing ability to self-propel in various ways responding to external field stimuli. The fascinating part is that they efficiently consume energy from the stimuli and convert it to mechanical energy leading to swimming towards particular directions. Studies of these stimuli-responsive architectures spurred the development of artificial counterparts in the form of micro- and nano-machinery for targeted bio-medical applications.

Herein, a new class of artificial microswimmers is reported that exhibit strong and complex phototactic response covering wide spectrum of light. The Colloidal Carbon Microswimmers (CCMSs) resembling green algae swimming, exhibiting directional propulsion and light mediated reversible colony formation (Figure 1; Supplementary video S1-S3). Being an excellent absorber and reflector of light, the assembly of the microswimmers carry the momentum of the light, inducing self-propelled motion by converting it to mechanical energy. During swimming process they interact with the other microswimmers of various architectures and form colonies throughout the observed media. These colonies are fractal in nature and highly dynamic and capable of collective locomotion (supplementary video S1-S3) responding to light of various wavelengths [20-21]. A fully organic colloidal nanostructure having extensive surface charge coagulates by charge-balancing via attachment and exhibit properties of green algae. Unlike the naturally occurring green-algae which exhibit photo-tactic motion and colony formation in aqueous solution, the Colloidal Carbon Microswimmers (CCMSs) replicate these phenomena in alcohol media, hence are called "alco-algae" (Figure 1).

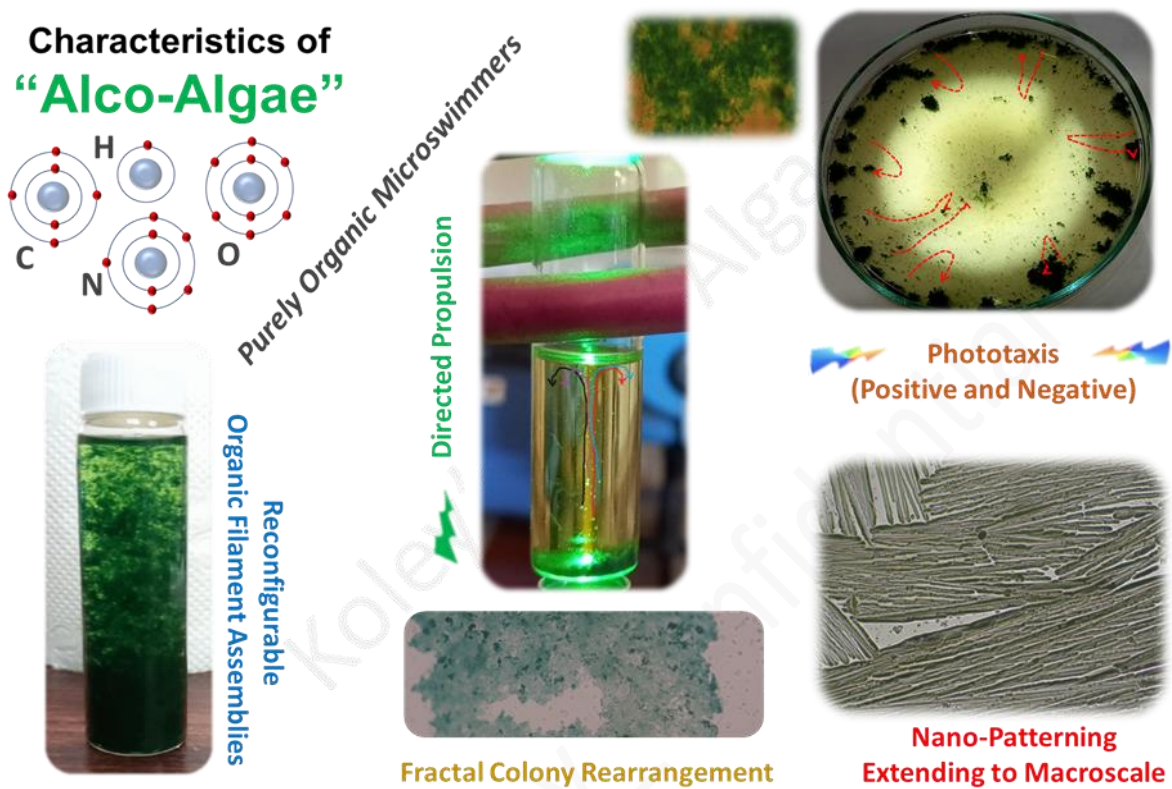

**Figure 1. General Characteristics and versatility of alco-algae microswimmers.** Schematically illustrated: A complex chemical compound composed simple and abundant of organic elements lead to preparation of green filamentous assemblies floating and swimming in alcoholic solutions. Multimodal though unique light matter interactions are observed akin to green-algae behaviour in aqueous environment. The alco-algae absorbs light with excellent efficiency and tend to form colonies under light exposure. A strong pointed light exposure causes rapid directed propulsion, whereas moderate intensity light gradient endorses phototactic motion with both positive and negative types rearranging the dynamic fractal colonies. When solidified, the motile alco-algae changes its form to exhibit Thallus like patterning resembling leaf like and diatom like morphologies. The multitude collective behaviour resembling algae life in an artificial systems endorse the nomenclature.

These alco-algae are made of fully organic compounds from the precursor comprising of carbon, nitrogen, oxygen and hydrogen. Carbon is one of the most abundant and active element remaining key source of organic life taking part in any information or matter transport in a living system. It forms complex rigorously with other abundant elements providing the transport chain and requisite architecture for biological functionality[22-24]. Here a synthesis route of nitrogen-doped carbon nanoparticles is exploited to achieve active colloidal nanocarbon repeating units that later take part in exciting phenomena resembling micro-algae (Supplementary information S1)`. In alcohol, the CCMSs produce larger scale (extending to cm dimension) microswimmers via fractal colony formation[25-26]. The collective swimming of dark green masses that can be visually observed. When a dilute solution of the CCMSs in alcohol is viewed under microscope, fibrous green masses of various sizes with lively movements are seen (Supplementary video S4). The architectures of micron onwards are seen to possess dynamic arms at the surface attached to a central body providing helical and twisted motion of the CCMSs (Supplementary video S5). Small dynamic branching units are highly charged and propel with simultaneous rotational and translational motion in all directions freely under homogeneous light source (Supplementary information S2). During swimming, these branched green architectures attach with other swimmers and swim together, or get attached to a relatively stable colony (Supplementary video S2, S6).

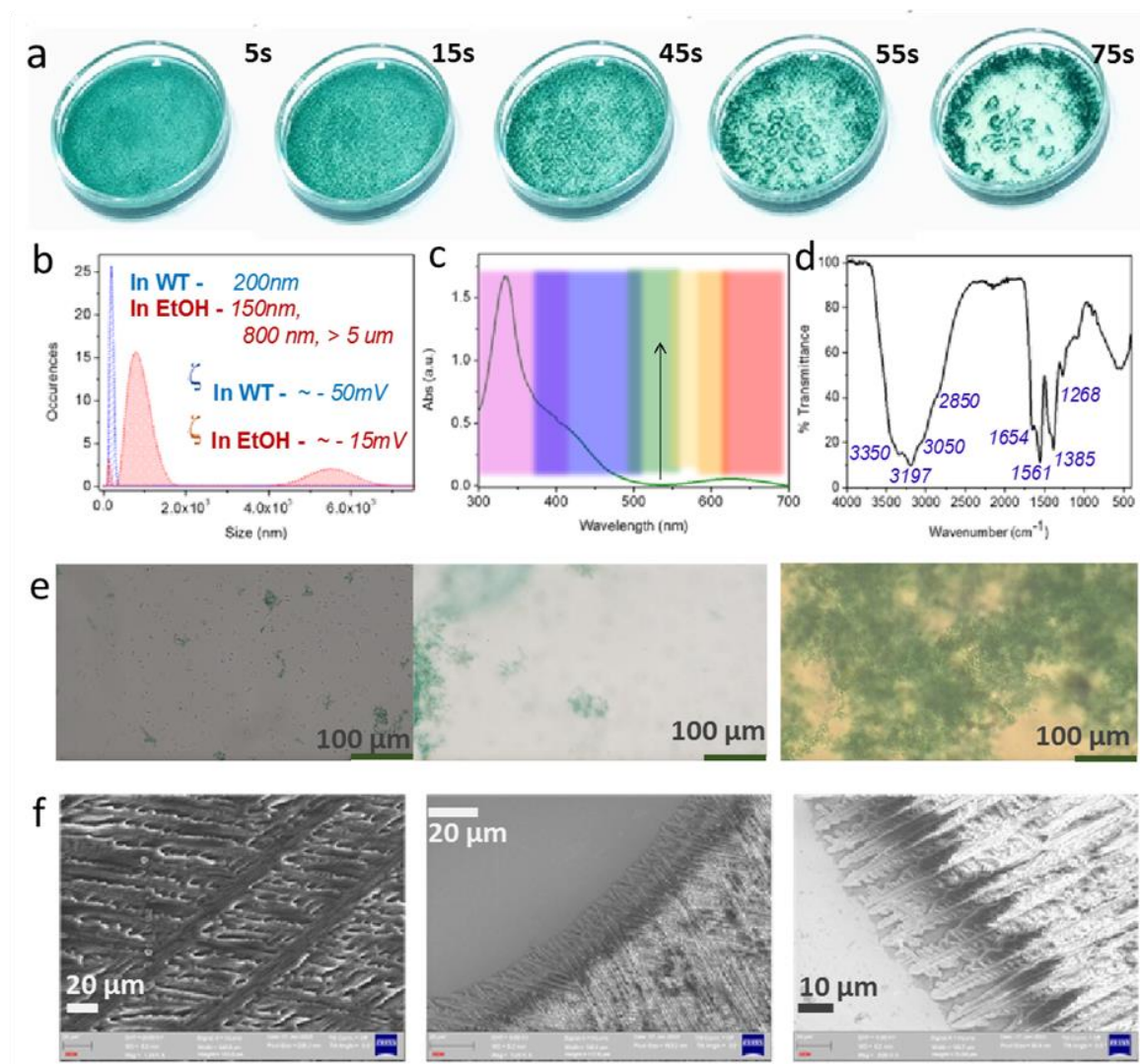

**Figure 2. Details of formation, optical, and structural properties of microswimmers alco-algae.** (a) From a cleaned and dense solution of CCMS pigment, when exposed to excess alcohol, swims vigorously and rapidly join together to form local colonies until achieve a giant structures. (b) Dynamic Light scattering (DLS) measurement of Hydrodynamic size of the CCMSs. (c) Absorption spectra of CCMS alco-algae reveal extensive absorption of ultraviolet and visible light. This compound absorbs all colours of UV-Vis light except the green. (d) FTIR spectra of CCMS reveal the presence of nitrogen in the carbon backbone and amide linkages along with free N-H and hydrogen bonded –OH groups in the surface. (e) optical micrographs of the swimming alco-algae; at low concentration and weak light exposure helical motion is seen; upon providing more light intensity, the motion becomes faster, tend to form colonies for energy dissipation; at high concentration giant colonies are formed in presence of moderately strong light exposure.

(f) Scanning electron microscopy images of the dried alco-algae ensemble. A highly planner morphology with weaving patterns of various highly organized nanostructures extending to micro to macro region is seen.

The formation mechanism of CCMSs relies on a fundamental colloidal rule of lyophilic sols[27-28]. These green mass is separated and when is examined via solvent tuning it is found to be consisting of lyophilic sols exhibiting super-hydrophilic characteristics. The Furrier Transform Infrared (FTIR) spectrum reveals the extensive presence of hydrogen bonded hydroxyl and amine groups making the surface readily to be solvated in water (Figure 2). Dynamic Light Scattering (DLS) experiments reveal 200nm hydrodynamic size of the CCMSs in water. These stable colloidal sols scatter light showing Tyndall Effect (supplementary information S3). These hydrophilic sols are covered with a layer of charged functional groups possessing -50 mV zeta potential value. In case the sol particles are composed of lyophobic sols with a protective lyophilic layer, they can undergo coagulation with a treatment of suitable solvent but in a reversible manner. Alcoholic solution does this magic, the addition of alcohol in the digested hydrophilic gel (Supplementary Text S3) readily starts to coagulate forming green colonies. In alcohol solution, the ~200nm population decreases drastically to form at first ~800nm sized sols and subsequently micron and millimetre sized colonies. During this process the surface charge is balanced and a far less value of -15mV with a narrow distribution is obtained with zeta potential measurements (Figure 2).

The origin of green colour from these alco-algae follows plant-like mechanism. Figure 3 exhibits the absorption spectra of a set of CCMSs in water. Considerable light absorption with strong bands are seen in the UV to blue region (300nm-490nm; peaks at 335nm and 420nm) and a moderate intensity broad band is seen in yellow to red region (~570nm - 680nm; peak at 625nm). In heteroatom mixed carbon nanostructures the UV absorption bands are assigned to $\pi$-$\pi$* transition of C=C bond and visible absorption bands are assigned to n-$\pi$* transition of C=O and C=N bonds[23]. Since the material absorbs all the component of light except the green, the CCMSs appear green in the solution while white light is passed through them. Yet they respond efficiently to the green light exhibiting directional propulsion. The alco-algae ensemble formed with the CCMSs possess highly patterned fractal structure[29]. Scanning Electron Microscopy (SEM) images reveal a long-range continuous pattern of nanostructures in dried self-assembled films. While a weaving like pattern is observed in the middle, the terminal of the patterns shows spikes. Groups of highly ordered dense-patterned mass and highly branched sparse-patterned mass are observed. A variation in the distribution of patterns are seen which is possibly responsible for various types of colonies depicted in optical micrographs (Figure 2e). A variation in the distribution of patterns are seen which is possibly responsible for various types of colonies depicted in optical micrographs. The body of the microswimmers possess highly ordered planner morphology, and the spike-like terminals endorse further attachment with other suitable terminals. Irregular anisotropic nature of the branching units derives dynamic local colonies of various architectures. The elemental composition study with Energy dispersed X-ray analysis of the patterned microstructures reveal the material on an average contains ~40-45% Carbon, ~30-35% Nitrogen and 20-25% Oxygen. FTIR studies revealed the presence of C=N, amide N-H, H-bonded -OH bonds within the system, which leads to a graphitic carbon nitride base with highly polar functional groups ready to involve in H-bonding. The CCMS compound corresponds to roughly a composite of N-doped carbon backbone providing basic architecture, amide bonds providing flexibility during motion, and extremely reactive and dynamic polar surface groups providing fractal nature.

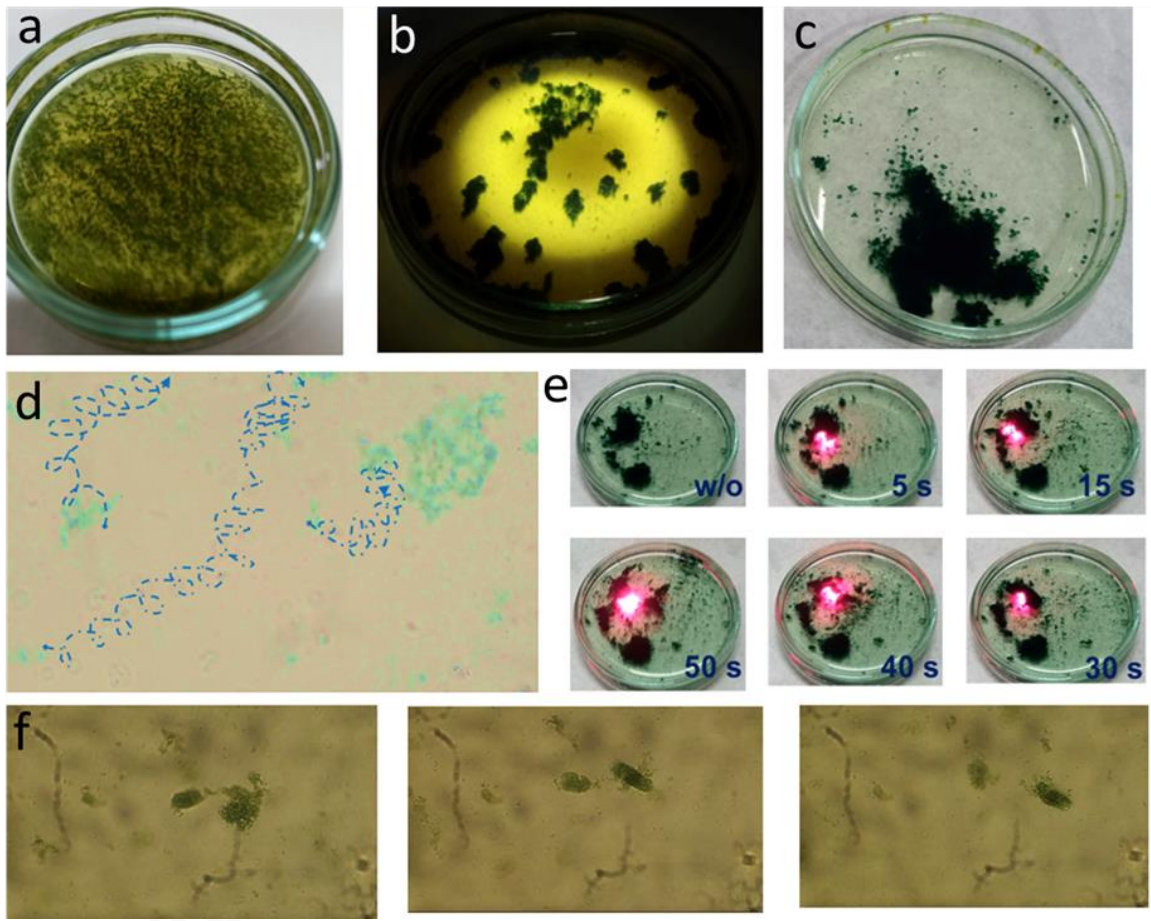

**Figure 3. Multitude fractal assemblies of alco-algae microswimmers in alcoholic solution.** (a-c) Various types of assemblies formed under (a) room light, (b) gradient light exposure, (c) in dark. (d) Helical motion of the motile microswimmers. The large sized colonies >100 micron are less motile, settle down though accept smaller motile microswimmers to form larger colonies. (e) Break-downing a stable group of colonies with laser light exposure. Strong pointed light source with concentrated energy can be efficiently absorbed and successively larger conies are divided into smaller fractal colonies depending upon the area and time of exposure. (f) During swimming a slow swimmer of ~50 micron sized colonial architecture get attached to a relatively larger more stable colony forming a flexible weak bond. Both undergo structural change while connected. They detach and start swimming again when an intruder swimmer pass through the weak bonds between them. Hence colonial swimmers dynamically attach and detach to dissipate energy and to rearrange their colony in the form of a better swimmers thus harvesting the light energy.

The primary direction of movement for artificial microswimmers is governed by the stimuli gradient applied. However, the navigation of life-like microswimmers in a fluid media is complex due to presence of local fluctuations and long-range fluid mediated factors induced by hydrodynamic interaction with the surface, interface, and obstacles[30-32]. Figure 3 demonstrate the macroscopic depiction of behaviour for a group of microswimmers colony. Creation of light gradient induces positive phototaxis leading to departing CCMS from a colony in shade and a summersault like swimming in the lighted area returning to the dim area joining a neighbouring colony (Supplementary video S2). While viewed microscopically, all types of rearrangements viz. rotational and translational motion, attachment and detachment, fractal dissolution of alco-algae colonies are observed to happen in moderate light intensity. However when the light intensity is high, the microswimmers create a turbulent flow in the solution rather than performing colony dynamics (Supplementary information S4; Video S7-S8).

The colonies are never static under in homogeneous light exposure. The swimming speed of the microswimmers varies with size, shape, light intensity and nearby environment. Structure of the submicron are not seen yet seen as they reflect light limited by the resolution. Few micron sized clusters are smallest observed and found to be most quick in swimming and joining to other swimmers in solution. The medium sized (10-50 micron) elongated CCMSs with branches at both the ends perform twisted rotation movement and follow helical pathways of swimming. After this size regime, the fractal assemblies become heavy, movement become sluggish and tend to stable colony formation down to the solid surface. Thus various types of phototactic motions with variable speed are seen in an alco-algae ensemble. The interaction with the patterned assembly closely matches to behavior of a class of green algae and hence artificial leaf.

These CMSs while highly concentrated form gel and when dried, forms layers of patterned structure mimicking the morphology of artificial leaves (Figure 4). Drying and change in morphology of the CMSs are dependent on the humidity level owing to strong hydrophilicity of the CMS units (Supplementary information S4). The self-assemblies exhibit fascinating macroscopic patterning, possess defined microscopic details[33-35]. These dried leaf like self-assembly transmit and reflect green light when white light passed through them. The absorption spectra of the fundamental units in water shows strong and defined absorption peaks at the blue and red part of the visible spectrum and thus appear as green when white light is passed through them. In an assembled condition a significant portion of green light is also absorbed owing to the patterned and layered nature of the microscopic details akin to a natural leaf.

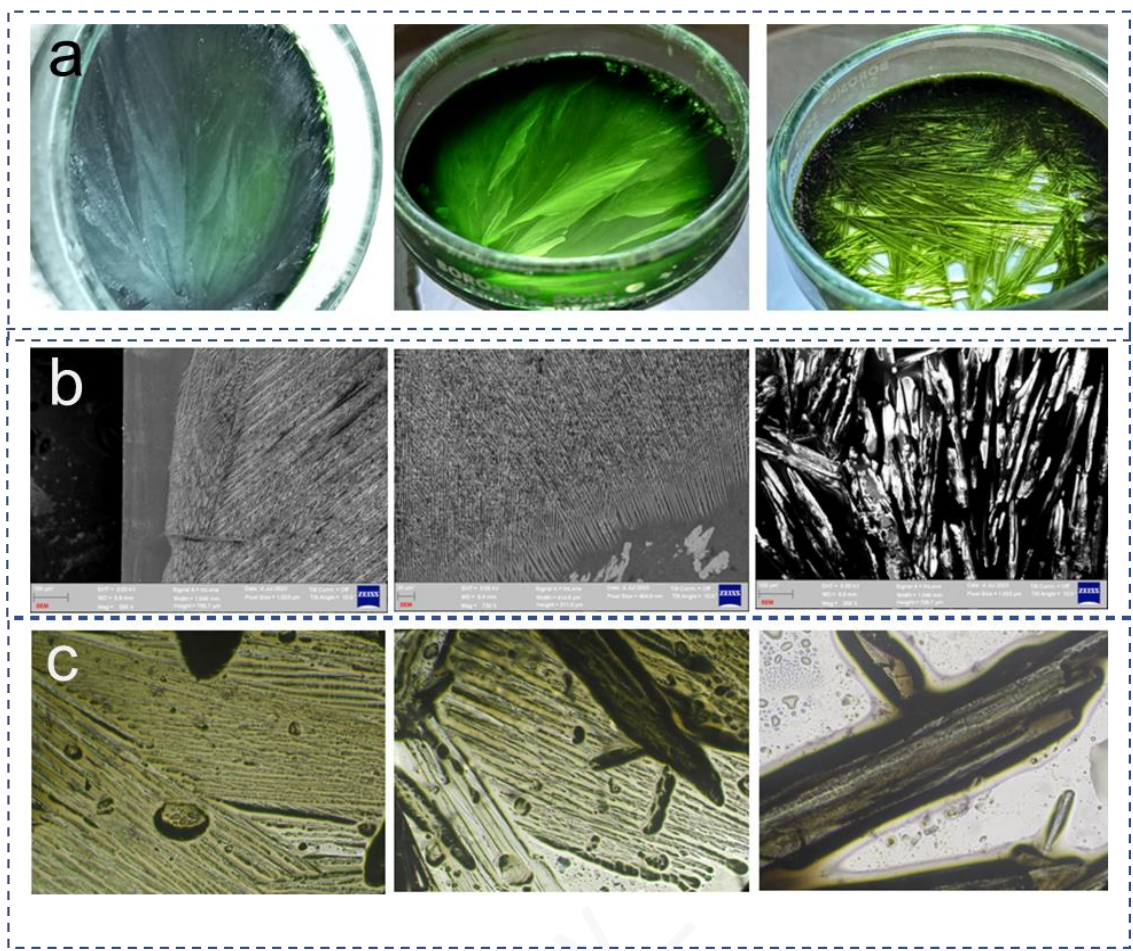

**Figure 4. Patterning Characteristics (nanoscale patterning extending to macroscales) of condensed alco-algae ensemble.** Top-Panel: Visual version photographs of condensed alco-algae in a Petri-dish with room-light and white light transmission. Bright-green leafy and layered patterning are observed. Mid-Panel: SEM images of various patterned condensed alco-algae architectures. Bottom-Panel: Optical micrographs of patterned leafy and needle architectures exhibiting layered channels of green chromophores. Diatom like small structures are also seen along with the main connected patterns.

Thus the systematic evolution of an organic compound exhibit complex light-matter and matter-matter interaction and a highly correlative patterning resembling natural life. The chemical processing remain the key to achieve life-like chemical assemblies from naturally fundamental and abundant atomic species. Hot organic atomic precursors form polar hetero-atom doped nano-carbon compound [36]. The successful processing follows a slow thermodynamically favourable pathway which provides time to the chemical bodies to be digested enough to build basic colloidal architecture ready to undergo fractal arrangements with hydrogen bonding. This study tracts the multitude possible interactions in the highly energetic chemical bodies in solution and condensed phase. The alco-algae compounds lose their motility in condensed phase and aqueous solution, only in alcohol media, active colloids are generated having ability to swim and interact (Physically and electronically) with the surroundings. The compound can successfully carry the momentum of light directionally and possess defined electrochemical potential (Supplementary information S5). Being completely organic and super hydrophilic, the cytotoxicity and environmental concerns are less worrisome. This endorses the utility of alco-algae swimmers in cargo-delivery, remote-catalysis, and other biomedical and agricultural application[37-39]. Most importantly these types of microswimmers can be generated from common and non-expensive chemical setup and the effects of vigorous interactions are very visual. We experience multitude observations of naturally occurring micro-organisms simultaneously performing bio-chemical process of various types harvesting light energy. The driving force remain self-sustaining which is special aspect of life. Naturally elements are processed under different ambient conditions and chemical exposure over a long period of time which lead to highly organized intelligent chemical systems performing energy absorption and conversion. With the organic life like active compound formation, if we do targeted mimic, we can understand the factors affecting the organic life at least from the chemical aspect.

**Acknowledgements**

S.K. thanks INSPIRE program of Department of Science and Technology (DST-INSPIRE), Govt of India for research support through a Faculty Fellowship and research grant. K.K.N. thanks Apex Project of Department of Atomic Energy, Govt. of India. We cordially thank Dr. Subhadip Ghosh, NISER Bhubaneswar for the generous support of laboratory and instrumental facilities access. S.K. thanks Mr. Debopam Acharjee, and Mr. Mrinal Kanti Panda for their assistance in spectroscopic measurements at the common instrumental facilities at NISER Bhubaneswar. We thank Dr. Sachindra Nath Sarangi for the assistance in laboratory, Mr. Soumya Ranjan Mohanty, and Dr. Paramita Maiti, Institute of Physics for their assistance in performing electron microscopy.

**Author Information**

Dr. Somnath Koley (Corresponding author): DST-INSPIRE Faculty, Institute of Physics, Bhubaneswar. ORCID: 0000-0002-6826-1501.

Prof. Karuna Kar Nanda: Director, Institute of Physics, Bhubaneswar; Professor, Material Research Centre, Institute of Science Bangalore. ORCID: 0000-0001-9496-1408

**Competing interests**

The authors declare no competing interests.